\documentclass{article}\usepackage{jkas2}
\runningauthor{BLASI}
\runningtitle{Cosmic Ray Acceleration}
\beginpage{1}
\endpage{5}

\begin{document}
\font\twelvei = cmmi10 scaled\magstep1 
       \font\teni = cmmi10 \font\seveni = cmmi7
\font\mbf = cmmib10 scaled\magstep1
       \font\mbfs = cmmib10 \font\mbfss = cmmib10 scaled 833
\font\msybf = cmbsy10 scaled\magstep1
       \font\msybfs = cmbsy10 \font\msybfss = cmbsy10 scaled 833
\textfont1 = \twelvei
       \scriptfont1 = \twelvei \scriptscriptfont1 = \teni
       \def\mit{\fam1 }
\textfont9 = \mbf
       \scriptfont9 = \mbfs \scriptscriptfont9 = \mbfss
       \def\bmit{\fam9 }
\textfont10 = \msybf
       \scriptfont10 = \msybfs \scriptscriptfont10 = \msybfss
       \def\bmsy{\fam10 }

\def\etal{{\it et al.~}}
\def\eg{{\it e.g.,~}}
\def\ie{{\it i.e.,~}}
\def\lsim{\raise0.3ex\hbox{$<$}\kern-0.75em{\lower0.65ex\hbox{$\sim$}}}
\def\gsim{\raise0.3ex\hbox{$>$}\kern-0.75em{\lower0.65ex\hbox{$\sim$}}}

\title{Cosmic Ray Acceleration during Large Scale Structure 
Formation}

\author{Pasquale Blasi}
\address{INAF/Osservatorio Astrofisico di Arcetri\\
Largo E. Fermi, 5 50125 Firenze (Italy)}

%


\address{\normalsize{\it (Received October 31, 2004; 
Accepted December 1,2004)}}

\abstract{
Clusters of galaxies are storage rooms of cosmic rays. They confine 
the hadronic component of cosmic rays over cosmological time scales 
due to diffusion, and the electron component due to energy losses. 
Hadronic cosmic rays can be accelerated during the process of 
structure formation, because of the supersonic motion of gas in
the potential wells created by dark matter. At the shock waves
that result from this motion, charged particles can be energized
through the first order Fermi process. After discussing the most 
important evidences for non-thermal phenomena in large scale structures,
we describe in some detail the main issues related to the acceleration 
of particles at these shock waves, emphasizing the possible role of 
the dynamical backreaction of the accelerated particles on the plasmas 
involved. 
}

\maketitle

\section{Introduction}

The non-thermal pressure support in galaxy clusters is currently 
unknown and historically considered as negligible compared with
the thermal pressure contributed by the material fallen in the 
gravitational potential well of dark matter and thereby heated to
approximately the virial temperature. On the other hand, the
discovery of the confinement of cosmic rays in clusters 
(Volk et al. 1996, Berezinsky, Blasi, \& Ptuskin 1997)
implied that the energy content of these structures in the form 
of cosmic rays should increase with time. It is therefore a natural 
conclusion that, if cosmic rays are efficiently accelerated in the
cluster neighborhood, they will be confined therein for cosmological
time scales and eventually represent some sizeable fraction of the
total pressure support. The inelastic proton-proton interactions of
these cosmic rays with the intracluster gas are the most direct 
channel for the production of gamma rays, through the generation 
and decay of neutral pions, and in fact current gamma ray observations
have already been used to constrain the cosmic ray energy density 
in a limited set of clusters, for different assumptions on the 
spatial distribution of the non-thermal pressure (Blasi 1999). 
Additional limits have been obtained more recently from the comparison
of the newly measured flux of radio emission from the Coma cluster 
at a few GHz (Thierbach et al. 2003) and the predicted radio emission 
from secondary electrons produced in pp scatterings (Reimer et al. 2004). 
Most of these limits are rather model dependent and sensitive to 
the spatial distribution and energy spectrum of cosmic rays in the 
intracluster medium. 

More recently it has become clear that the role of electrons accelerated
at shocks arisen during the formation of the large scale structure of
the universe is crucial for the generation of gamma radiation (Loeb \& 
Waxman 2000) because of the relatively short lifetime of these particles 
for inverse Compton scattering (ICS) off the photons of the cosmic microwave 
background (CMB). This emission is expected to contribute to an uncertain 
extent to the extragalactic diffuse gamma ray background (EDGRB): in 
(Loeb \& Waxman 2000)
it was argued that all of the EDGRB could be explained in terms of ICS
of relativistic electrons, while later estimates led to numbers one
order of magnitude lower (Gabici \& Blasi 2003b). Contrary to some 
statements appeared in the literature (Colafrancesco 2002; Kawasaky \& Totani 
2002), no association of clusters of galaxies
with unidentified EGRET gamma ray sources could be assessed in a 
statistically significant way (Reimer et al. 2003; Scharf \& Mukherjee 2003).

Despite the absence of firm proof of gamma ray emission from clusters
of galaxies or more in general from large scale structures, strong 
evidences exist now of the presence of non-thermal phenomena in this
class of objects. Extended radio emission in the form of radio halos
and radio relics have been found (see (Feretti et al., 2004) and references 
therein for a review of these phenomena), while some clusters of galaxies 
also appear to be sources of hard X-ray emission (see (Petrosian, 2004) and
references therein for a review).
There are hints of a correlation between the presence of radio halos
in rich clusters of galaxies and the occurrence of a recent merger
event (Buote, 2001). 

The models that attempt to explain the observations fall in three 
classes, that we refer to as {\it primary} electron models 
(Fujita \& Sarazin 2001, Blasi 2001, Miniati et al. 2001a), 
{\it secondary} electron models (Dennison 1980, Colafrancesco \&
Blasi 1998, Blasi \& Colafrancesco 1999, Dolag \& Ensslin 2000), 
and {\it reacceleration} models (Schlickeiser, Sievers \& Thiemann 1987,
Brunetti et al. 2001, Petrosian 2001, Brunetti et al. 2004).
The first two are 
distinguished depending on whether the radiating electrons are directly 
accelerated through some kind of mechanism, or rather generated as a 
result of production and decay of charged pions in inelastic proton-proton
scatterings. 
In the reacceleration models the radiating electrons are re-energized 
through the interaction of MHD waves with a population of relic electrons, 
pre-existing the formation of the radio halo. 

Present observations, mainly those in the radio band, seem to suggest
that the extended diffuse radio halos are best explained by the 
reacceleration models, while many arguments can be identified against
the primary and secondary electron models, as discussed below
and by Brunetti in these proceedings (Brunetti 2004).

This paper is structured as follows: in section II we describe
the main attempts to explain the observations and the diagnostic 
tools that can be used to discriminate among them. In section
III  we describe our view of the perspective for detection of
gamma rays from clusters of galaxies. In section IV the process 
of particle acceleration at shock waves created during the formation 
of clusters of galaxies and filaments is described. We show there 
that the backreaction of the accelerated particles on the shock 
itself is likely to be strong. We conclude in section V.

\section{Non-thermal phenomena in clusters and their explanation}

The richest set of data on non-thermal activity in clusters is related
to radio halos (Feretti et al. 2004) and the best studied radio halo is that
of the Coma cluster, despite the fact that it is certainly not one of
the brightest. Observations show a roughly power law spectrum extending
to a few GHz with a steepening that appears to be a cutoff, as confirmed
by the recent measurement of Thierbach et al. (2003). The surface brightness of
the Coma cluster has also been measured at different frequencies: the
extension of the radio halo has been shown to be smaller at high frequencies
than it is at low frequencies. In the inner part of the cluster, the 
spectrum of the radio emission is as flat as $\nu^{-0.8}$ (Giovannini et al. 
1993). This spectral
steepening as a function of the distance from the center of the cluster
is one of the most challenging pieces of information to be explained.
This wealth of information on the radio emission contrasts with the 
poorness of the data on the hard X-ray excess, which appears to be 
present only in a few clusters (Petrosian 2004). The models that attempt to 
explain the observations fall in three classes, that we will refer to
as {\it primary} electron models, {\it secondary} electron models, and
{\it reacceleration} models. The first two are distinguished depending 
on whether the radiating electrons are directly accelerated through 
some kind of mechanism, or rather generated as a result of production 
and decay of charged pions in inelastic proton-proton  
scatterings. In the reacceleration models the radiating electrons are
re-energized through the interaction of MHD waves with a population of
relic electrons, pre-existing the formation of the radio halo. 

In the following we will concentrate our attention on the extended radio
emission rather than on the X-ray excess, since it provides a much richer
set of constraints on the physics of the acceleration and propagation of
cosmic rays in the intracluster medium. 

The {\it primary} models are those in which electrons are directly 
accelerated at shock waves formed during merger events or other 
violent processes in the cluster volume (Fujita \& Sarazin 2001,
blasi 2001,Miniati et al. 2001a).
While these models can easily reproduce the shape of the volume-integrated
spectrum of the radio radiation, as generated through synchrotron emission
of electrons with an {\it ad hoc} spectrum, the morphology of
the generated radio halo is unlike that observed in radio halos: the 
short lifetime of the required electrons, due to ICS energy losses, 
implies that the emission should be concentrated in rim-like regions
around the sites where the electrons are accelerated (for instance
the merger related shock waves), while the observed radio emission 
appears to be spread out over Mpc scale structures. In addition to 
the extended emission, radio radiation is sometimes also observed in 
the vicinity of shock waves (Markevitch 2004, these proceedings). 

In the inner part of the clusters, where the radio halos are brighter,
the merger related shock waves have very low Mach numbers (Gabici \&
Blasi 2003a)
and the spectra of accelerated particles are therefore too steep to
explain observations (similar results were obtained by Berrington \&
Dermer (2003) and by Inoue (these proceedings)).
Moreover, the efficiency of acceleration at these shocks is expected to 
be very low (see for instance Fig. 6 of (Ryu et al. 2003) 
and Fig. 3b here), although this expectation depends somewhat 
on the injection recipe assumed for the calculations. 
On this basis it appears unlikely that electrons accelerated as primaries
at merger related shocks can explain the spectrum and morphology of the
observed radio halos. 

In this perspective, {\it secondary electron models} appear to be very 
appealing: in these models, electrons are generated through the decays of 
charged pions, resulting from inelastic proton-proton interactions. The 
confinement of cosmic rays within the cluster volume (Volk et al .1996,
Berezinsky et al. 1997)
over cosmological time scales enhances the energy density in the form 
of cosmic rays in the intracluster medium and this increases the probability
for these particles to interact and produce secondaries. It is worth
reminding that the protons confined in the intracluster medium also 
include most of the cosmic rays accelerated outside the clusters and
eventually advected inside through the accretion flow, driven by gravity.
This accretion process results in the formation of shock waves with 
potentially very high Mach numbers, since they propagate in a cold 
non-virialized medium (Bertschinger 1985). These shocks were considered as
sites for particle acceleration by Volk et al. (1996) and by Berezinsky
et al. (1997), while clear evidence
is found of spatially extended filamentary-like shock surfaces in numerical 
simulations (Ryu et al. 2003). Protons accelerated at these shock waves would 
finally end up being accumulated within clusters of galaxies, and the
spectrum of these cosmic rays is expected to be as flat as $E^{-2}$, 
as predicted by the linear theory of particle acceleration. This expectation
is satisfied only as far as the fraction of the energy dissipated by the
shock is very small, while non-linear effects modify this prediction
in a substantial way when an appreciable fraction (even of the order of
$10\%$) of the kinetic energy crossing the shock is converted into 
accelerated particles. Note that even in the strongly non-linear regime, 
the fraction of particles which are accelerated can be very small, of 
the order of $\sim 10^{-4}$. 

Secondary electron models were first proposed by Dennison (1980)
and considered in detail by Colafrancesco \& Blasi (1998) and by 
Blasi \& Colafrancesco (1999). More recently these models
have been revived by many authors (e.g. Dolag \& Ensslin (2000) and
Miniati et al. 2001b) for radio halos and (Pfrommer \& Ensslin 2004) 
for radio mini-halos). It has been known for some
time now that the very general features of the observed radio halos 
could be reproduced by secondary electron models (see for instance 
(Blasi 2001) for a calculation involving both primary and secondary
electron models). What is hard or impossible to explain through these
models are the details, which therefore become the tool to discriminate 
among different explanations. 

The maximum energy of accelerated protons depends on many unknown 
quantities, and on the specific acceleration process responsible
for the injection of protons. Within clusters, as discussed in 
(Berezinsky et al. 1997), there are many potential sources of cosmic rays:
normal galaxies are expected to provide only a small fraction of 
the cosmic rays in the intracluster medium; active galaxies and 
shock waves associated with the process of structure formation
are expected to be the dominant sources of cosmic rays in large 
scale structures. In all these cases the estimates of the maximum 
energy of the accelerated particles are large enough that no 
cutoff in the GHz region of the synchrotron emission of the 
secondary electrons should be expected. Secondary electrons 
generate power law radio spectra that extend way beyond the 
GHz region. This point was recently used in (Reimer et al. 2004) 
to impose a limit on the energy density in the form of cosmic rays in 
the intracluster medium of the Coma cluster. 

The spectrum of the synchrotron emission in the context of the 
secondary electron model is expected to be rather independent of
the spatial location in the cluster, so that no spectral steepening
is expected, while the observations show it, at least in the case of
the Coma cluster and in some other cases in which spatially
resolved spectra are available. Both these issues point strongly 
against the secondary electron models. One could play with the 
maximum energy of the protons and try to obtain the spectral 
steepening as well as the steepening in the volume integrated 
spectrum, by lowering this maximum energy way below the existing
estimates. However, even assuming that this procedure has any physical
motivation, it was shown by Brunetti (2003) and more recently in 
(Brunetti 2004) that the energetic requirements for the model 
to work are unacceptably large and therefore to be discarded
on the basis of the observations listed above. In addition, if
the evidence (Buote 2001) that seems to be emerging of a correlation 
between recent or ongoing mergers and the presence of radio halos is 
confirmed by future observations, the secondary electron model 
has an additional problem in that the radio emission would be 
dominated at any time by the electrons produced by the pile up 
of cosmic ray protons during the merger history of the cluster, 
rather than by the last merger event. No correlation should 
therefore be expected. It is worth stressing once more that all
these conclusions have been derived on the basis of observations
carried out in very few clusters. As such, these conclusions
are hardly extendable to a more general context at the present
time.

The third class of models of the non-thermal activity in the 
intracluster medium is that based on the possibility that low
energy electrons, relics of the past activity of the cluster
or injected in a past flare event from an active galaxy in the
intracluster medium, could be re-energized due to resonant or
non-resonant interactions of these electrons with MHD turbulence,
possibly associated with the merger history of the cluster.
These models were first introduced by Schlickeiser, Sievers \& Thiemann 
(1987), and later investigated in great details by Brunetti et al. 2001,
and by Petrosian (2001).

The nice feature about these models is that electrons develop 
a so-called inverse spectrum, with a bump at high energy. This
bumpy structure cuts off at the maximum Lorentz factor of the
electrons which is typically of the order of $\sim 10^5$, as
determined by the balance between the acceleration rate and 
the rate of energy losses for ICS. The maximum energy is relatively
low since the acceleration process is alike a second order 
Fermi process and is therefore rather inefficient compared with
shock acceleration. The presence of a bump in the spectrum 
implies that the synchrotron emission at a given frequency is 
dominated by electrons with different energy (and therefore
with different spectrum) depending upon the local strength of 
the magnetic field. This process easily accounts for the spectral
steepening, while the volume integrated spectral cutoff is 
naturally produced by the presence of the maximum Lorentz 
factor. The time during which the process is effective is
of only a few hundred million years, so that the emission is 
expected to correlate with the most recent or even ongoing merger 
event, and gradually dies out with time after the merger
(Brunetti et al. 2004). 

The rather disappointing aspect of this explanation is
the complexity of the mechanism: despite the basic physics
involved is rather simple, the details of the development 
of the MHD turbulence and its interaction with the relic
electrons is all but understood. Each type of MHD turbulence
has its own channels of wave-particle interactions and there
are numerous different types of turbulent modes that can 
be excited (Alfven waves, slow and fast magnetosonic modes,
Whistler modes, and so on). When fluid turbulence is injected in 
the intracluster medium, the mechanism for its conversion 
to MHD turbulence is not established as yet, although several 
possibilities have been investigated: one of the channels which
are used most often in the recent literature is the so-called
Lighthill mechanism (Fujita, Takizawa \& Sarazin 2003), 
which couples the fluid and MHD turbulence 
and therefore allows to carry out detailed calculations of the
development of the MHD waves and their interaction with charged
particles. The cascading process from large to small spatial 
scales, those most relevant for the acceleration process (at least 
for Alfven turbulence), is
also matter of much investigation in plasma physics, but it would
be an underestimate of the complexity of the situation to say
that it is by now clear how this non-linear process works. 

All these different pieces of the model need to be further 
investigated and probably much input can come from the study 
of turbulence in environments in which we have more control, 
such as in the solar system or near the Earth or 
again in the Galaxy, where second order Fermi processes
are also expected to be at work. Despite the complexity 
of the mechanism however, the basic features of the observations
can easily be accomodated within the {\it reacceleration models}, 
which makes them unique at the present time. 

Until recently, the calculations involved in the reacceleration 
model were carried out in a stationary regime, namely assuming that
both the spectra of electrons and waves were time-independent.
Moreover there was the hidden assumption that electrons were the 
only population of charged particles present in the intracluster 
medium. These assumptions were both relaxed in a recent work 
(Brunetti et al. 2004), 
where the fully time-dependent coupled equations for electrons,
protons and waves were solved, in the case of Alfv\`en waves as
the main channel of MHD turbulence. The presence of the protons
(thermal and non-thermal) is crucial for understanding the effects
of the interactions of electrons with Alfv\`en waves, because
these waves can resonate with thermal protons and relativistic
protons, while they only resonate with suprathermal electrons. 
The net effect of the presence of the protons is to reduce the
transfer of energy from waves to electrons, and therefore 
damp the acceleration process. It was calculated that if more 
than $\sim5\%$ of the thermal energy of the cluster is in the form
of relativistic protons with a power law spectrum, the radio halos 
would not be effectively generated (Brunetti et al. 2004). 
This constraint can be 
substantially relaxed in the case of magnetosonic waves (Cassano 
and Brunetti, in preparation).
More work is being carried out on the role of turbulent reacceleration
on the secondary electrons and positrons generated from pp collisions
(Brunetti and Blasi, in preparation), which represents a realistic 
situation in which the radio emission would be generated by reaccelerated 
secondary electrons, but the gamma ray signal may still be present 
because of a substantial presence of hadrons.

\section{Gamma Rays from Clusters of Galaxies}

The main motivation for interest in the gamma ray emission arose
as a consequence of the possibility of cosmic ray confinement in
the intracluster medium, and consequently the possibility that 
radio halos could be the result of synchrotron emission from 
secondary electrons. A natural by-product of the model is the
copius production of gamma radiation as a result of the decay 
of neutral pions. Calculations of this emission from single clusters 
of galaxies have been carried out by Ensslin et al. (1997), Blasi 
\& Colafrancesco (1998) and Blasi (1999).

Gamma radiation in the intracluster medium can also be generated 
as a result of the ICS of high energy electrons off the universal 
photon background. This finding was proposed by Loeb \& Waxman (2000) as a
possible explanation of the extragalactic diffuse gamma ray 
background. In (Gabici \& Blasi 2003b) it was shown that the contribution of
clusters of galaxies to the diffuse gamma ray background was 
actually limited to $\sim 10\%$ of the measurement reported by
Sreekumar et al. (1998) (see (Strong, Moskalenko \& Reimer 2004) 
for a revised estimate of the
extragalactic gamma ray background). This contribution is 
dominated by the accretion process, rather than by the 
acceleration of electrons during merger events, as a consequence
of the weakness of the typical shocks involved in merger events in 
the central parts of clusters (Gabici \& Blasi 2003a). This new estimate
appears to be close to the results reported in (Keshet et al. 2004).

\begin {figure*}[t]
\vskip -0.6cm
\centerline{\epsfysize=5.8cm\epsfbox{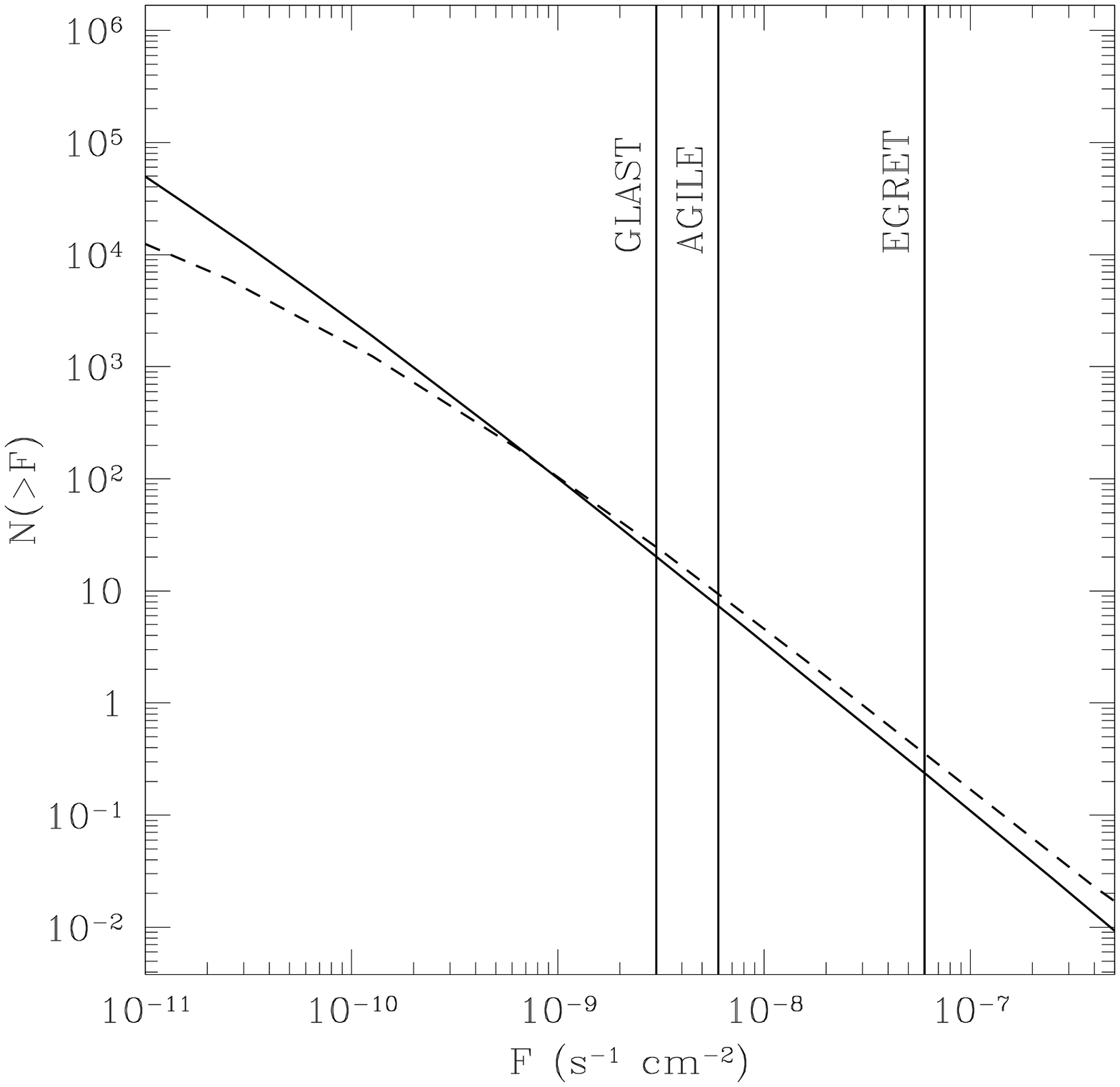}
\epsfysize=5.8cm\epsfbox{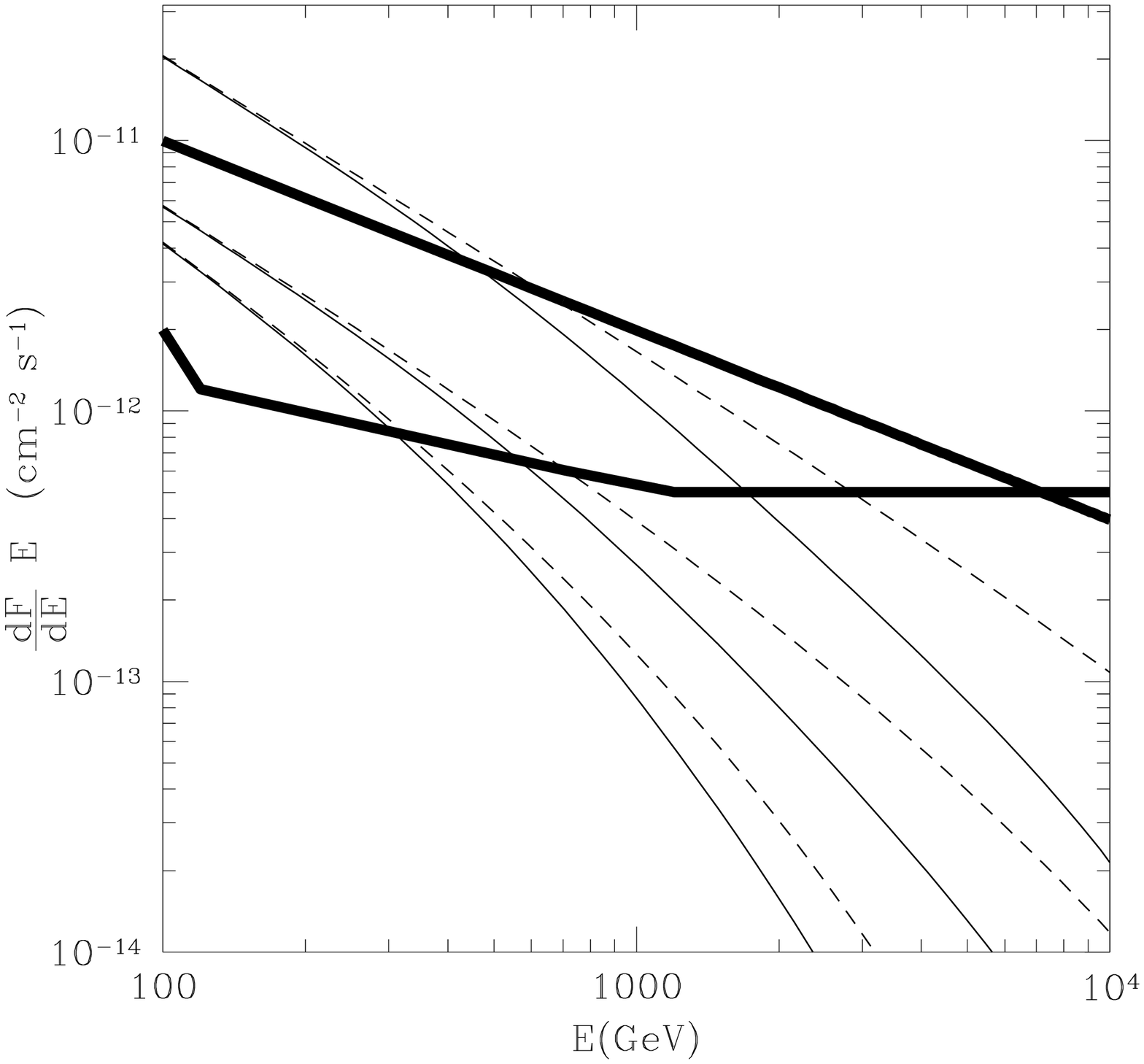}}
\caption{a) Number of accreting (solid line) and merging (dashed line) 
clusters with gamma ray flux greater than $F$. The vertical lines 
represent the GLAST, AGILE and EGRET sensitivity for point sources. b)
Gamma ray emission in the 100 GeV - 10 TeV region. The thick solid
lines represent the sensitivities of a IACT for point sources (lower curve) 
and extended sources (upper curve). The predicted gamma ray fluxes from a 
Coma-like cluster at a distance of 100 Mpc with and without absorption of 
the infrared background are plotted as dashed and solid lines respectively.
}
\vskip -0.5cm
\end{figure*}

The detectability of the gamma ray signal from the electrons 
accelerated at either merger or accretion related shocks 
also depends on the strength of the shocks formed in either 
one of these cases. In (Gabici \& Blasi 2004) the LogN-LogS of clusters 
of galaxies as gamma ray sources was calculated, as it is shown
in Fig. 1a, where the typical sensitivities of GLAST, AGILE
and EGRET are also indicated. The calculations refer to the
case of a constant efficiency of electron acceleration of
$5\%$, independent of the Mach number of the shocks. 
The electrons accelerated in this way provide a negligible 
contribution to the diffuse radio emission and to the hard
X-ray emission. 

Although the initial motivation to believe that there may
be a gamma ray emission from clusters of galaxies was 
related to the need to explain the non-thermal activity
of such clusters at other wavelengths, the present situation
appears to be quite different: secondary electron models 
appear to be inadequate to properly describe observations,
at least in those few cases in which the data are good enough,
as for the Coma cluster. As a consequence, in those cases
the present observations can only be used to impose limits on the
amount of non thermal protons confined in the intracluster
medium, and therefore on their gamma ray signal. 
In addition to this possible pp signal, as illustrated in Fig. 1a, 
GLAST should be able to identify the ICS signal, if the efficiency 
for particle acceleration of electrons is large enough. 

The corresponding signal in the TeV region is much more uncertain:
the decay of neutral pions may provide a flux in this energy 
region, as shown for instance by Blasi (1999). However the
energy density in the form of cosmic rays required for a 
measurable flux is close to or even in excess of the equipartition 
level. As for the ICS signal from electrons accelerated at
shocks during structure formation, the maximum energy and the
efficiency for electron acceleration are the crucial quantities.
Assuming that the magnetic field in the acceleration region
is large enough (or amplified enough) to allow for the acceleration
up to multi-TeV energies, then the ICS emission may extend up
to a few TeV. 

In Fig. 1b we plot the results of our calculations for the
high energy gamma ray spectrum generated from a Coma-like cluster of galaxies 
at 100 Mpc distance for the case of merger and accretion. The effect of the
gamma ray absorption in the infrared background (IRB) is illustrated by 
the difference between the solid lines (with absorption) and dashed lines 
(without absorption).

From top to bottom, the lines refer to three different cases: 1) a merger
between two clusters with masses $10^{15}M_\odot$ and $10^{13}M_\odot$; 2)
an accreting cluster with mass $10^{15}M_\odot$ with a magnetic field
at the shock in the upstream region $0.1\mu G$; 3) an accreting cluster with 
mass $10^{15}M_\odot$ with a magnetic field at the shock in the upstream 
region $0.01\mu G$. 

The thick solid lines represent the sensitivities for a generic Cherenkov 
telescope as calculated in (Aharonian et al. 1997). These results are obtained 
considering an array of imaging atmospheric Cherenkov telescopes (IACT) 
consisting of $n$ cells, each consisting of a $100\times 100~\rm m^2$ 
quadrangle with four '100 GeV' class IACTs in its corners. The two thick 
curves in Fig. 1b represent the minimum detectable fluxes 
for point sources (lower curve) and an extended $1^{\circ}$ wide source 
(upper curve) for an exposure time of 1000 hours. The exposure time here 
is defined as the product between the observation time and the number 
of cells that form the array. For instance, an exposure of 1000 hours 
can be achieved with a 100 hours observation performed by an array 
consisting of 10 cells.

\section{Shock Acceleration during Structure Formation}

The supersonic motion of baryon-loaded dark matter clumps results 
in the formation of shock waves, if there is enough magnetic field 
in the background medium to mediate the build-up of a collisionless
shock. At least in the case of mergers between clusters, these shocks
are indeed observed (Markevitch, these proceedings). More problematic 
is the detection of shock 
waves in the intergalactic medium outside the virialized regions, 
since the gaseous environment is much less dense and the temperatures
are much lower there. On the other hand these external shocks are
crucial for understanding the role of cosmic rays in large scale
structures, because most cosmic rays accelerated in these outskirts
are finally advected in the central parts, where they get confined 
for cosmological times (Volk et al. 1996, Berezinsky et al. 1997). 

In recent times, much interest has arisen on the distribution of 
Mach numbers of the shock waves related to structure formation,
since the spectrum of the accelerated particles is fully determined
by the Mach number, at least in the context of the linear theory 
of shock acceleration. In (Miniati et al. 2000) a histogram was
presented of the number of shocks per unit Mach number in the 
central Mpc of a cluster, which showed a pronounced peak at Mach 
number $M=5$, while practically no shock was {\it detected} at 
lower values of $M$. A simple model for the development of merger 
related shocks in the hierarchical picture of structure formation
(Gabici \& Blasi 2003a) showed that in fact a pronounced peak at $M\sim 1.5$
was to be expected. This conclusion appears to be confirmed, at 
least qualitatively, even from observations: for all cases in 
which the Mach number of merger related shocks could be measured,
the inferred Mach number is $M\sim 1-3$ (Markevitch, these proceedings).

It is sometimes argued that in the approach of Gabici \& Blasi (2003a)
strong shocks are not correctly accounted for. However, strong shocks 
are associated very seldomly with merger events, and more often with 
phenomena occurring in the outskirts of the cluster. These shocks 
were included in (Gabici \& Blasi 2003a, 2004) where the accretion 
processes were accounted for. 

The fact that simulations were missing the weak
shocks was confirmed by the work of Ryu et al. (2003), in which 
simulations were carried out for different choices of 
the spatial resolution in the simulations. That work showed
that: 1) an increasing number of weak shocks was {\it visible} 
when increasing the resolution; 2) most energy was dissipated 
at Mach numbers $M\sim 2-4$. Both conclusions have been confirmed
by more recent simulations by the same authors. 

The second finding actually hides several complex pieces of 
information: in (Ryu et al. 2003), the dissipation in the form of 
cosmic rays was evaluated by adopting a simple but physically
motivated recipe for the injection, namely the so-called 
{\it thermal leakage injection}. This recipe adopts an injection momentum 
$p_{inj}$ as a multiple of the momentum of thermal particles $p_{th}$ 
and assumes that the particles that take part in the acceleration
process are those with momenta in excess of $p_{inj}$. The choice 
of $p_{inj}$ is a guess, since the details of particle injection
still represents one of the most serious problems of particle
acceleration at shock waves. With the reasonable choice made by 
Ryu et al. (2003), the shocks that dissipate most energy in the form 
of cosmic rays are those with Mach numbers $M\sim 2-4$. 
Unfortunately, in this narrow range of Mach numbers, the spectra
of accelerated particles change from very steep ($E^{-3.3}$) to 
very flat ($\sim E^{-2.26}$) in the linear theory of shock acceleration,
so that it becomes crucial to understand which shocks take most of
the energy.
The question is therefore {\it what is the role of weak shocks?}
Weak shocks can certainly contribute to heat the gas in the 
intracluster medium, but they also play an important role in
reaccelerating the pre-existing cosmic rays. The role of re-acceleration 
was considered by Gabici \& Blasi (2003a) but is ignored in other calculations.
Moreover, their role in the acceleration of fresh particles from
the thermal pool depends very sensibly on the recipe for the injection
(Blasi, Gabici and Vannoni, in preparation), and this recipe is all but 
established: the conclusion that shocks with Mach number $\sim 2$ are 
inefficient accelerators depends on this recipe and should not be taken 
as a model independent and universally true statement.

The most important recent developments in the study of particle 
acceleration at newtonian shock waves are related to the effects
of the backreaction of the accelerated particles upon the shock
itself. For a strong shock, linear theory implies that the particle 
spectrum approaches $E^{-2}$, which is energy divergent, unless 
a maximum momentum is adopted. Even doing so, it can happen that 
the pressure in the form of cosmic rays can approach or exceed the 
available kinetic energy $\rho u^2$. As a consequence, the 
backreaction of the accelerated particles cannot be neglected
any longer. Moreover, the escape of the particles at the maximum
momentum make the shock {\it radiative} and therefore more 
compressible. All these effects go in the direction of making
the shock a more efficient accelerator, therefore implying a 
stronger backreaction. This is a typical example of a non-linear
run-away system. The main effects of this non-linear backreaction are
1) the production of a cosmic ray mediated precursor in the upstream
section; 2) the acceleration of particles to non-power-law spectra;
3) reduced efficiency of the shock in the heating of the background
plasma. 

The implications of these effects for large scale structures were
suggested by Kang (2003), Gabici \& Blasi (2004a) and Kang \& Jones 
(2004) and are now recognized as crucial
to understand the role of cosmic rays in clusters of galaxies and
their neighborhood. 

Several approaches, both analytical (Malkov 1997, Malkov, Diamond \&
Volk 2000, Blasi 2002, Blasi 2004) and numerical (Ellison 1991, Kang,
Jones \& Gieseler 2002) exist in the literature to treat 
the non-linear effects of particle acceleration. In order to 
illustrate the main results, we adopt here the approach first 
introduced by Blasi (2002), and generalized by Blasi (2004),
while adding the recipe of thermal leakage for the injection
(see (Gabici and Blasi 2004a) for further details).
This calculation allows us to calculate analytically and in a 
relatively simple way the spectrum of the accelerated particles, 
the shape of the precursor and the temperature of the gas at
any point upstream and downstream. The details of the calculations
will be published elsewhere (Blasi and Gabici, in preparation).

\begin {figure*}[t]
\vskip -0.6cm
\centerline{\epsfysize=5.8cm\epsfbox{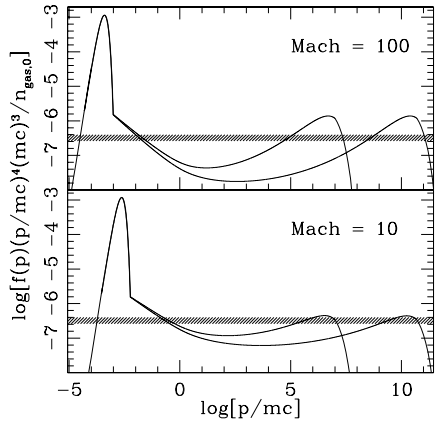}}
\caption{Proton spectra at the shock location for different values of 
the shock Mach number and of the maximum momentum of the accelerated 
particles.}
\vskip -0.5cm
\end{figure*}

In the calculations, the injection is chosen in such a way that all 
the particles with momentum larger than $p_{inj}=\xi p_{th}(T_2)$,
with $\xi=3.5$ are injected in the accelerator. It is worth stressing 
that the thermal momentum is fixed by the temperature of the downstream
gas, which is not a free parameter but an output of the non-linear 
calculations. In this way, the system regulates the amount of energy
to be transferred to the non-thermal particles. The only parameter 
left free is the maximum momentum $p_{max}$ of the accelerated particles 
which is determined by either the finite size of the shock or by 
energy losses. 

In Fig. 2 we plot the spectra of protons at the shock location for 
Mach number $M=100$ (upper panel) and $M=10$ (lower panel) and for 
$p_{max}/(m_c c) = 10^7$ and $p_{max}/(m_c c) = 5\times 10^{10}$. 
The normalization of the curves 
is such that these spectra reproduce those expected for protons 
accelerated at the accretion shock of a Coma-like cluster and therein
confined (see (Gabici and Blasi 2004a)). The concave
shape of the spectra, typical of non-linear particle acceleration at
shocks, is clear in the figures. It is also worth stressing that in current 
numerical simulations it is difficult to achieve maximum momenta larger 
than $\sim 100$ GeV (Kang \& Jones 2004), while for the application to 
clusters of galaxies much larger values are expected. In this respect, 
analytical calculations represent a unique tool to tackle the problem 
of evaluating the non-linear backreaction of cosmic rays on the shocks in 
large scale structures. The maxwellian distribution in Fig. 2 represents the 
spectrum of the particles in the downstream fluid, at the temperature 
determined through the calculations. 

As stressed above, one of the consequences of the non-linear backreaction 
of cosmic rays on the shocks induced by large scale structures is that
the gas is heated less than it would be at shocks where cosmic rays are
absent (if there were anything like that). In Fig. 3a, we plot the 
temperature ratio between downstream infinity and upstream infinity
$T_2/T_1$ for an ordinary shock (dotted line) and for a cosmic ray
modified shock for $p_{max}=10^2,~10^6$ and $5\times 10^{10}$ GeV (from 
top to bottom) as a function of the Mach number of the shock. It is
possible to see that typically the heating is not affected appreciably
for low Mach number shocks, while it is considerably suppressed for
strongly modified shocks. 

\begin {figure*}[t]
\vskip -0.6cm
\centerline{\epsfysize=5.8cm\epsfbox{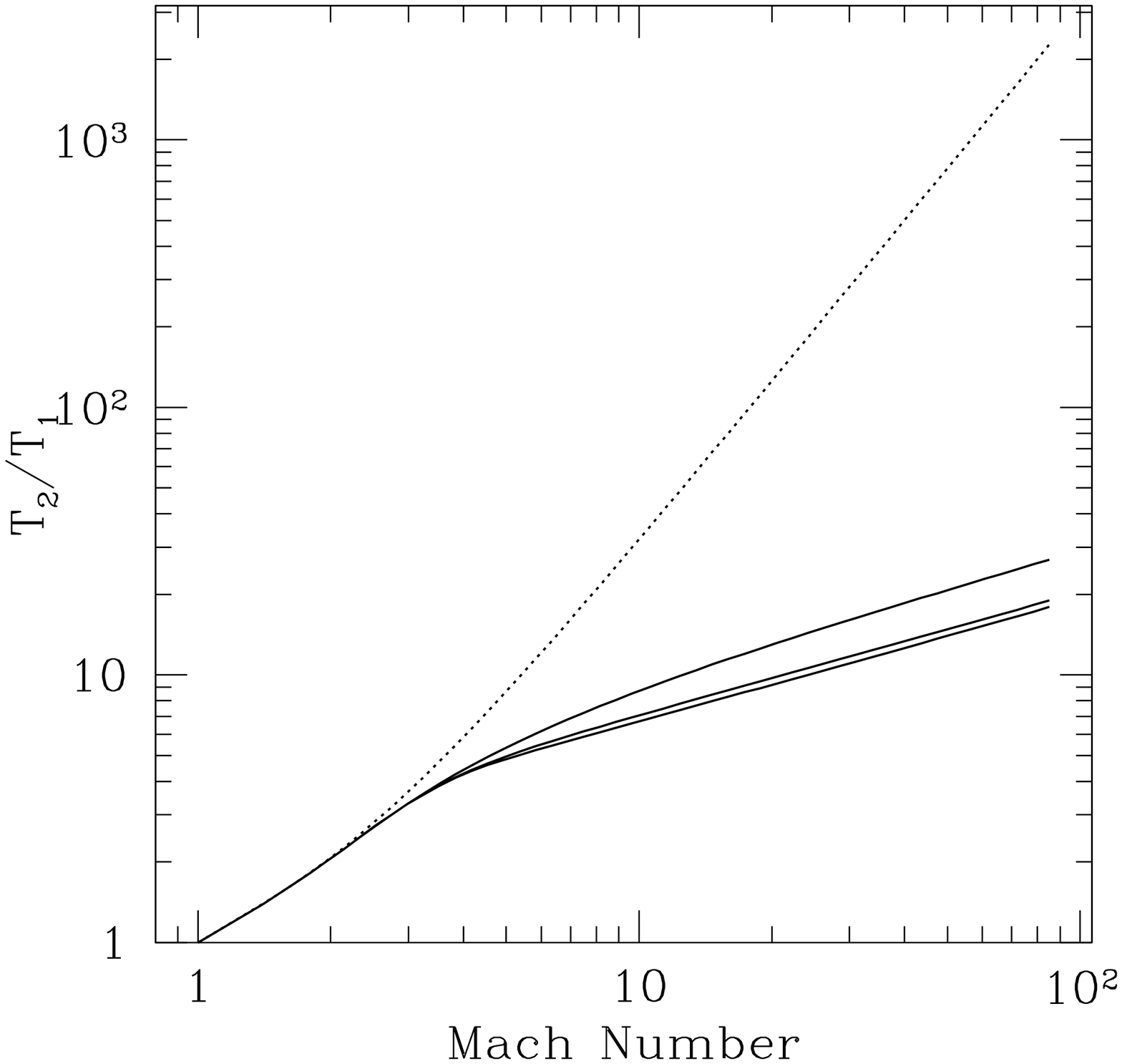}
\epsfysize=5.8cm\epsfbox{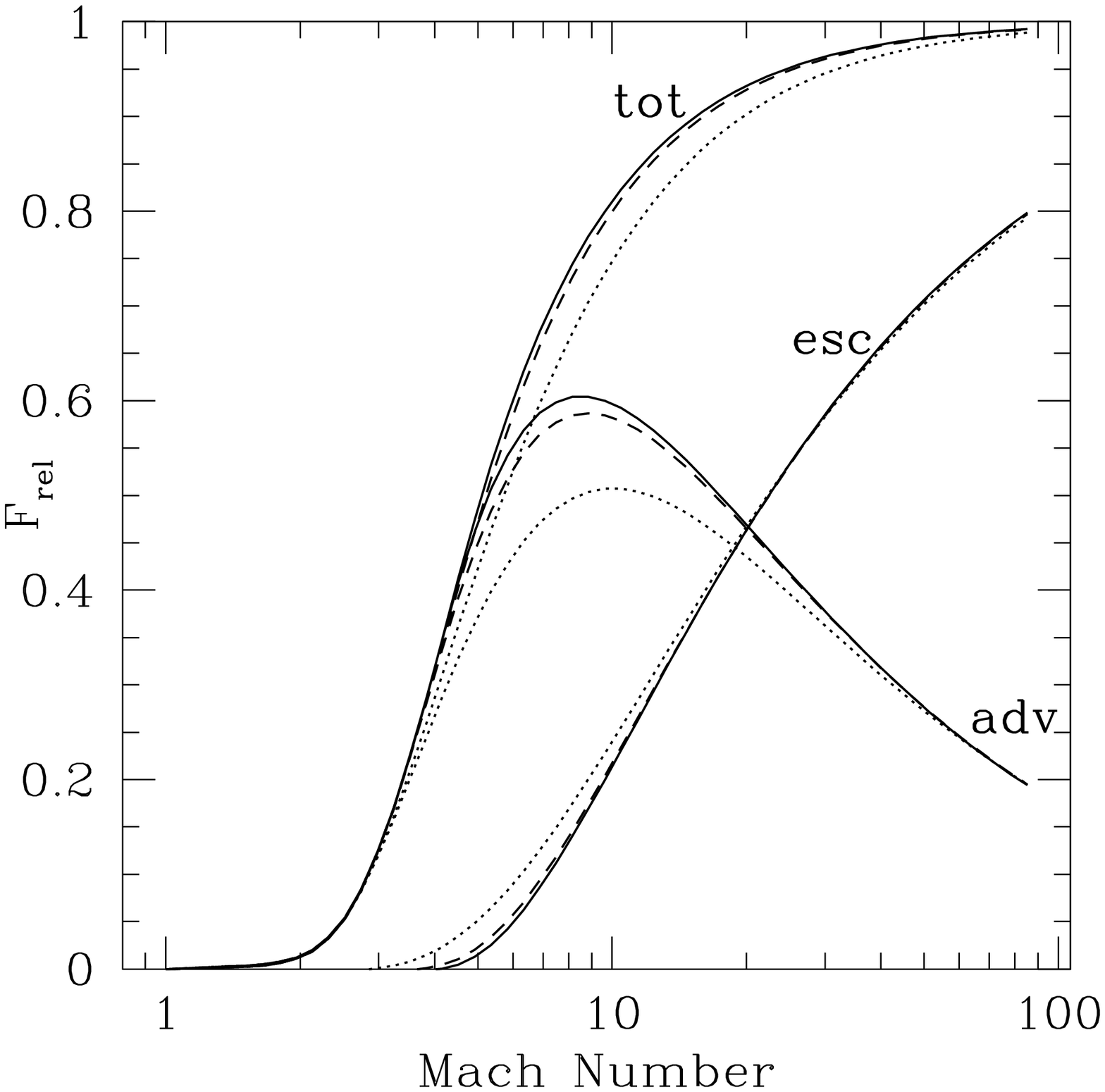}}
\caption{a) Temperature ratio between downstream and upstream infinity
as a function of the Mach number (see text). b) Efficiencies for advected 
particles, escaping particles and total efficiency as functions of the Mach 
number.}
\vskip -0.5cm
\end{figure*}
In Fig. 3b we also plot the fraction of flux (in units of 
$(1/2)\rho u^3$) which is effectively advected downstream, the 
fraction of flux which escapes at $p_{max}$ and the sum of the
two, which saturates to a number very close to unity for 
large Mach numbers. It is important to recognize that the 
relevant flux which is later confined in the cluster volume
is that advected downstream, rather than the total flux. 
The latter is in fact dominated for large Mach numbers by the 
escaping flux, which by definition does not end up at downstream 
infinity.

\section{Conclusions}

There is now clear evidence of the existence of non-thermal
phenomena associated with the formation of the large scale 
sctructure of the Universe. Most information on the mechanisms
responsible for the acceleration of the radiating particles
can be gathered from observations of the diffuse radio emission 
from the intracluster medium. 
In the few cases in which we have enough information about
the radio emission, as for the Coma cluster, the data on the
volume integrated spectrum, on the spectral steepening as 
a function of the distance from the center of the cluster,
and the correlation between radio halos and recent merger
events all seem to play against the so-called secondary
electron models. The morphology of these radio halos also
contradicts the possibility that the radiating electrons 
are accelerated as primary particles in merger related shock waves. 
The possibility that currently is favored by observations is that 
the radiating electrons are the result of reacceleration of relic 
electrons, possibly due to MHD turbulence. 

Shock acceleration at the collisionless shocks formed during
structure formation is a very important issue in the study 
of non-thermal phenomena in clusters of galaxies, as it has
already been recognized to be in the case for particles 
acceleration at shocks in supernova remnants. 
Indeed the typical Mach numbers and velocities of the 
shocks involved in the two cases are very similar. The 
confinement of cosmic rays in clusters might provide us with
a unique tool to explore the non-linear mechanism of 
shock acceleration (see for instance (Gabici \& Blasi 2004a) for an 
investigation of the generation of gamma rays in this regime).

\acknowledgements{
The author is very grateful to the organizing committee of the 
International Conference on Cosmic Rays and Magnetic Fields in 
Large Scale Structure for the funding support to attend the 
conference. He also thanks G. Brunetti, S. Gabici and G. Vannoni 
for the ongoing collaboration and useful discussions.}

\end{document}